\documentclass[twocolumn,aps]{revtex4}
\input{epsfig.sty}
\newcommand{\beq}{\begin{eqnarray}}
\newcommand{\eeq}{\end{eqnarray}}
\begin{document}
\title{Pulsar Kicks With Sterile Neutrinos and Landau Levels}
\author{Leonard S. Kisslinger\\
Department of Physics, Carnegie Mellon University, Pittsburgh, PA 15213\\ 
Ernest M. Henley\\
Department of Physics, University of Washington, Seattle, WA 98195\\
Mikkel B. Johnson \\
Los Alamos National Laboratory, Los Alamos, NM 87545 \\}

\begin{abstract} We use a model with two sterile neutrinos obtained by
fits to the MiniBoone and LSND experiments. Using formulations with
neutrinos created by URCA processes in a strong magnetic field, so
the lowest Landau level has a sizable probability, we find that with
known paramenters the asymmetric sterile neutrino emissivity might 
account for large pulsar kicks.

\end{abstract}
\maketitle
\noindent
PACS Indices:97.60.Bw,97.60.Gb,97.60.JD
\vspace{1mm}

\section{Introduction}

   The gravitational collapse of a massive star often leads to the formation
of a neutron star, a pulsar. It has been observed that many pulsars move with
linear velocities of 1000 km/s or greater. See  Ref.\cite{hp97} for a review. 
We have investigated the pulsar kicks which arise from the
modified URCA processes in the time interval 10-20 sec after the supernova
collapses, with a strong
magnetic field and temperature so that the population of the lowest Landau 
level is approximately 0.4 of the total occupation, and we find large pulsar 
kicks\cite{hjk07}.

   The largest neutrino emission after the supernova collapse takes place
during the first 10 seconds, with URCA processes dominant. The possibility
of pulsar kicks from anisotropic neutrino emission due to strong magnetic
fields during this time was discussed more than two decades ago\cite{chu84}.
It has been shown\cite{dor} that, with the strength of the magnetic field 
expected during this period, the lowest Landau level has a sizable 
occupation probability,
which produces the neutrino emission asymmetry that is needed for pulsar
kicks. However, due to the high opacity for standard model neutrinos in
the dense region within the neutrinosphere, few neutrinos are emitted, and
the pulsar kick is not obtained\cite{lq98}. Sterile neutrinos with a small 
mixing angle have small opacities.  It has been shown\cite{fkmp03}
that using the model of Ref.\cite{dor} and assuming the existence of a
heavy sterile neutrino (mass $>$ 1 kev), with mass and mixing angle
constrained to fit dark matter, the pulsar kicks might be explained.

  Recently, the MiniBooNE Collaboration found that the data for electron 
neutrino appearance showed an excess at low energies, in comparison to
what was expected in the standard model\cite{mini}. This data, along 
with the LSND data, has been analyzed in a model with two light sterile 
neutrinos\cite{ms07}, and compared to MiniBooNE data\cite{s07}. The
mixing angles of two light sterile neutrinos were extracted. See, however,
Ref.\cite{mm08}, which questions the accuracy of the results of 
Ref\cite{ms07}.

 In the present paper we use the fits of Refs.\cite{ms07,s07} with
two sterile neutrinos to investigate the possibility of obtaining
the large pulsar velocities which have been observed. As we shall show,
our model differs from that of Ref\cite{fkmp03} in that with a much larger
mixing angle there is a higher probability of sterile neutrinos, but
a much smaller effective volume, due to a larger opacity. However, as
we shall show, since the mean free path is much larger than those of
standard neutrinos, under the conditions in which standard neutrinos 
produce a pulsar velocity of 2-300 km/s, the MiniBoone/LSND sterile
neutrinos can give a kick of more than 1000 km/s.

\section{Asymmetric Sterile Neutrino Emissivity and Pulsar Kicks in Light
Two-Sterile Neutrino Model}

  Within about 1 second after the gravitational collapse of a large star,
the neutrinosphere with a radius of about 40 km, with temperature equilibrium,
is formed. For about 10 seconds about 98\% of neutrino emission occurs, with 
neutrinos produced mainly by URCA processes. Due to the strong magnetic field,
neutrino momentum asymmetry is produced within the neutrinosphere, but with
a small mean free path they are emitted only from a small surface 
layer of the neutrinosphere, and the pulsar kick cannot be accounted for.
If a standard active neutrino, say the electron neutrino, oscillates into
a sterile neutrino, it will escape from the protoneutrino star and 
neutrinosphere, unless it oscillates back into the active neutrino. The
mixing angle plays a key role. In the work of Fuller et al\cite{fkmp03}
the mixing angle is so small that the sterile neutrinos are emitted. In the
present work the starting point is the analysis of MiniBooNE and LSND data,
with the two or more sterile neutrinos with small masses and large mixing 
angles. Before we can proceed, however, it is essential to determine 
possible effects of the high density and temperature of the medium on the 
mixing angles.

\subsection{Mixing Angle in Neutrinosphere Matter}

   It has long been known that dense matter can effect neutrino states.
The famous MSW effect\cite{w78,ms85} for understanding solar neutrinos,
and the study of oscillations of high energy neutrinos\cite{ams05} are
studies of mixing of active neutrinos in matter. There have been many
other studies. In the present work we are dealing with sterile/active 
neutrino mixing  given by the mixing angle $\theta_m$ in neutrinosphere 
matter

\beq
\label{1}
      |\nu_1> &=& cos\theta_m |\nu_e> -sin\theta_m |\nu_s> \\
      |\nu_2> &=& sin\theta_m |\nu_e> +cos\theta_m |\nu_s> \; . 
 \nonumber 
\eeq

  In the work Ref\cite{fkmp03} it was shown that the mixing angle for sterile
neutrinos that can account for dark matter as well as those produced in the 
neutron star core is almost the same as the vacuum value. Starting from the 
much larger mixing angles for the sterile neutrinos that seem to account for
the MiniBoonE, LSND data, we need the value of the mixing angles in the
neutrinosphere, as we discuss below. The effective mixing angle in matter, 
$\theta_m$ can be related to the vacuum mixing angle, $\theta$ by\cite{afp01}
\beq
\label{2} 
          sin^2(2\theta_m) &=& \frac{sin^2(2\theta)}{sin^2(2\theta) +
(cos(2\theta)-\frac{2p V^T}{(\delta m)^2})^2} .
\eeq
In Eq(\ref{2}) $V^T$ is the finite temperature potential, while the
finite density potential due to asymmetries in weakly interacting particles
has been dropped as it vanishes when temperature equilibrium is 
reached\cite{afp01}. A convenient form for $V^T$, with the background of
both neutrinos and electrons included, is given in Ref\cite{nr88}
\beq
\label{3}
        V^T &=& \frac{28 \pi G_F^2}{45 \alpha} sin^2 \theta_W
(1+0.5 cos^2\theta_W) p T^4 \; ,
\eeq
with $G_F,\theta_W$ the standard weak interaction parameters and
$\alpha=1/137$. Assuming T=20$a$ MeV, with $a \leq$1.0, p=$b$ MeV, 
and $(\delta m)^2=$ 1.0 ev$^2$\cite{ms07,mm08}, we find
\beq
\label{4}
 \frac{2p V^T}{(\delta m)^2}& \simeq & 5.1 \times 10^{-3} b^2 a^4 \ll 
cos(2 \theta) \; .
\eeq

Therefore the mixing angle in the neutrinosphere medium is approximately 
the same the vacuum mixing angle. This agrees with Ref\cite{fkmp03}.

\subsection{Emissivity With a Light Sterile Neutrino}

   We now use the fits to MiniBooNe and LSND with light sterile neutrinos
to estimate pulsar kicks. The MiniBooNE results are consistent with the 
LSND results only if there are at least two sterile neutrinos. Models 
with three sterile neutrinos have also been considered\cite{sgmg,mm08}.
Fits to the MiniBooNE experiment and the LSND results by Ref.\cite{ms07}
in Ref.\cite{s07} with two sterile neutrinos are shown in Fig. 1.
\vspace{-.5 cm}

\begin{figure}[ht]
\begin{center}
\epsfig{file=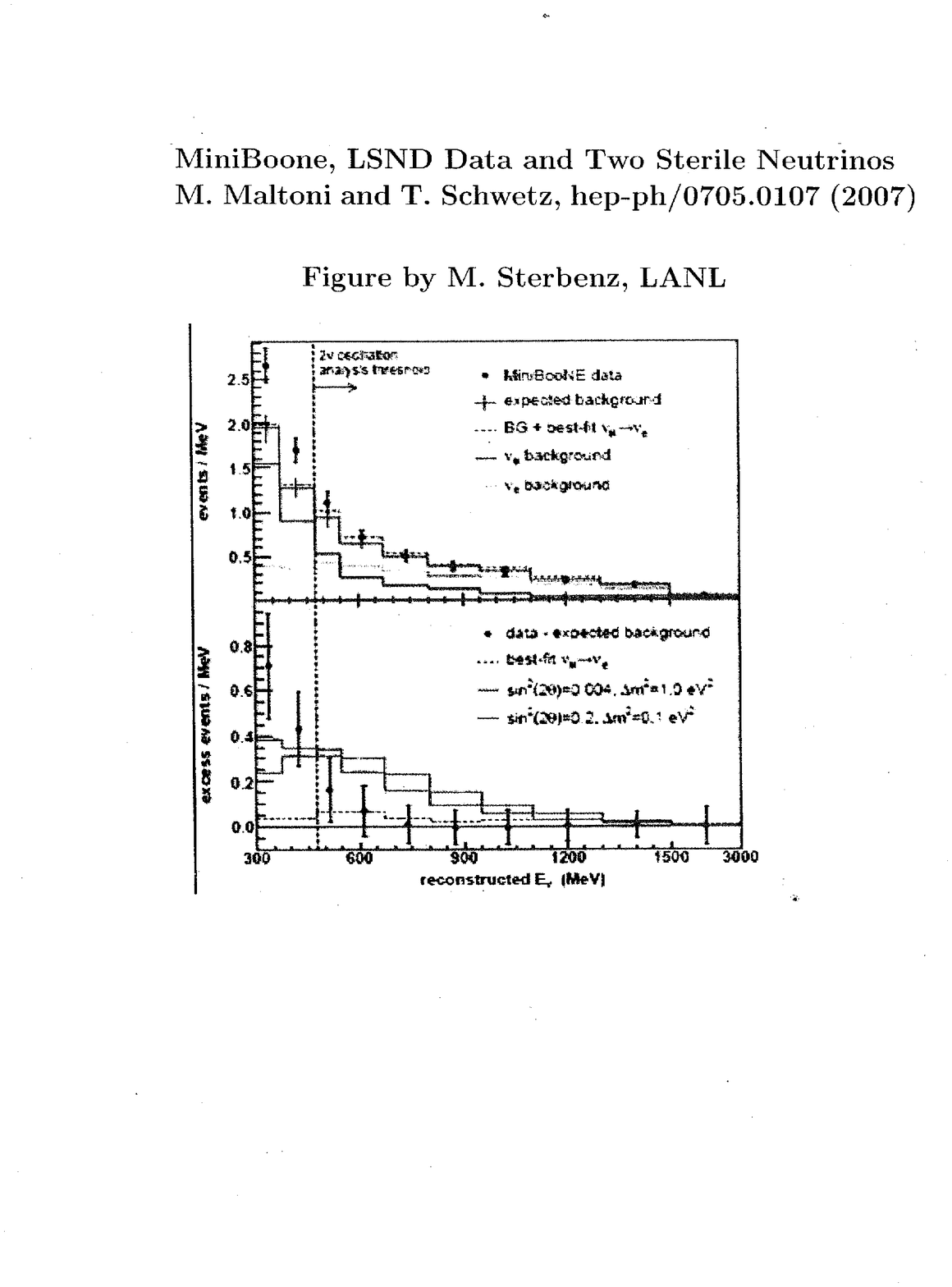,height=11.5cm,width=8cm}
\caption{$sin^2(2\theta_{1s})$=0.004; $sin^2(2\theta_{2s})$=0.2}
{\label{Fig.1}}
\end{center}
\end{figure}
\vspace{3mm}

From the Sterbenz/Maltoni-Schwetz fits one finds for the mixing angles of
the two sterile neutrinos:
\beq
\label{5}
                  (sin 2\theta_{1s})^2 &=& 0.004 \nonumber \\
                  (sin 2\theta_{2s})^2 &=& 0.2 \; ,
\eeq
and the masses are negligibly small. Note that this is in contrast to the
parameters of Ref.\cite{fkmp03}, with the constraint of dark matter giving
a mixing angle of $(sin 2\theta_{dm})^2 \simeq 10^{-8}$, and a mass greater
than 1 keV. Note that in a recent analysis by Maltoni~\cite{mm08} it was
found that the fit~\cite{ms07} to the MiniBooNE and LSND data had problems
with disappearance data, although there are rather large error bars. In our
present work we will use values for $(sin 2\theta)^2$ in the range 0.2 to
0.004 to estimate the pulsar kick.

The probability of asymmetric emission, giving a pulsar kick,
does not depend directly on the sterile neutrino mass in our model, but
is proportional to the $(sin 2\theta_s)^2$. It is the large mixing angles
found in fits to MiniBoone and LSND that lead us to carry out the
investigation in the present paper. 

\subsection{General Formulation of Neutrino Emissivity With a Strong Magetic
Field}
   The neutrino emissivity is given in general form in many 
papers, e.g., see Refs~\cite{fm,bw}:
\beq
\label{6}
   e^\nu &=&\Pi^4_{i=1} \int \frac{d^3p^i}{(2\pi)^3}
\frac{d^3 q^\nu}{2 \omega^\nu(2 \pi)^3} 
 \int\frac{d^3q^e}{(2\pi)^3}
\nonumber \\
 && (2\pi)^4 \sum_{s_i,s^\nu} \frac{1} {2\omega^e_L}  \omega^\nu \mathcal{F}
M^{\dagger} M \\
&&  \delta(E_{final}-E_{initial}) \delta(\vec p_{final}-
\vec p_{initial})  \nonumber \; ,
\eeq
where $M$ is the matrix element for the URCA process and
 $\mathcal{F}$ is the product of the initial and final Fermi-Dirac 
functions corresponding to the temperature and density of the medium.
The main source of the asymetric emissivity that produces the pulsar
velocity is the fact that the electron has a large probability to be
in the lowest (n=0) Landau level. See Refs~\cite{jl,mo} for a discussion 
of Landau levels. The asymetric emissivity can be seen by considering the
weak axial intereaction, $W_A$,
\beq
\label{7}
   W_A &=& -\frac{G}{\sqrt{2}}g_A\chi_p^\dagger \vec{l}\cdot \vec{\sigma}
\chi_n \\
           l_\mu &=& \bar{\Psi}(q^e) \gamma_\mu(1-\gamma_5)\Psi(q^\nu) 
\nonumber \; ,
\eeq
with $G=\frac{10^{-5}}{m_n^2},\;  g_A=1.26$, the $\chi$  are
the nucleon spinors, and the lepton wave functions are $\Psi(q^e),\Psi(q^\nu)$,
where  $q^e$ and $q^\nu$ are the electron and antineutrino momenta, 
respectively. The key to the asymmetric emission is given be the trace
over the leptonic currents, $Tr[l_i^{\dagger} l_j]$,
\beq
\label{8}
 \int d^2 q^e_\perp Tr[l_i^{\dagger} l_j] & \simeq & 8 \pi E^e [(q^\nu)^j 
\delta_{i3} +(q^\nu)^i \delta_{j3} \nonumber \\
  &&-\delta_{ij} (q^\nu)^3]  (\hat{q}^e = \hat{B} = \hat{z}) \; ,
\eeq
with the magnetic field B in the z direction. We only consider the weak
axial force, which is dominant. Using the relationship given in Eq(\ref{8})
one can show that the result of the traces and integrals over the axial 
product matrix element has the form ($\hat{B} = \hat{z}$)
\beq
\label{9}
   \int\int |M_A|^2 &\propto& (q^\nu)^z \; .
\eeq
Details are given in Ref~\cite{hjk07}, where it is shown that the asymmetric
neutrino emissivity, using the general formulation of Ref~\cite{fm}, is
\beq
\label{10}
  \epsilon^{AS} &\simeq& 0.64 \times 10^{21} T_9^7 P(0)\times f
\nonumber \\ 
&&{\rm erg\; cm^{-3}\; s^{-1}}=p_{ns} c \frac{1}{V_{eff} \Delta t} \; ,
\eeq
where $T_9 = T/(10^9 K)$, $p_{ns}$ is the neutron star momentum, $P(0)$ is 
the probability of the electron produced with the antineutrino
being in the lowest Landau state, f=.52 is the probability of the
neutrino being at the + z neutrinosphere surface~\cite{hjk07}, $V_{eff}$ is
the volume at the surface of the neutrinosphere from which neutrinos are
emitted, and $\Delta t \simeq 10 s$ is the time interval for the emission.
 
We derive $P(0)$ and $V_{eff}$, the effective volume for the emissivity, 
in the next two subsections.

\subsection{$P(0)$ = Probability for the Electron to be in the n=0 Landau 
Level}

Just as in our previous work in which Landau levels play a crucial 
role\cite{hjk07}, only the lowest Landau level, for which the helicity
is -1/2 (rather than $\pm$1/2 as with the usual Dirac spinors) gives
asymmetric emission. The probability that an electron in a strong magnetic
field is in the lowest (n=0) Landau level, $P(0)$, can be calculated from the
temperature, T, and the energy spectrum of Landau levels\cite{jl,mo}.
A particle with momentum p and effective mass $m_e^*$ in a magnetic field 
B in the nth Landau level has the energy

\beq
\label{11}
           E^L(p,n) &=&  \sqrt{p^2 +(m_e^*)^2 + 2 (m_e^*)^2 \frac{B}{B_c} n} 
\; ,
\eeq
with $B_c = 4\times 10^{13}$ G, and $m_e^*$ is the 
effective mass of the electron at the high density of the protoneutron 
star and neutrinosphere.  
 
 From standard thermodynamics the probability of occupation of the n=0
Landau level, P(0), is given by\cite{fkmp03}:
\beq
\label{12}
   P(0) &=& \frac{F(0)}{F(0) + 2 \sum_{1}^{\infty} F(n)} \; ,
\eeq
where F(n), with magnetic field B, temperature T, and chemical potential $\mu$,
is 
\beq
\label{13}
       F(n) &=& \int_{p_{min}}^{\infty} dp \frac{[m_n-m_p-E^L(p,n)]^2}
{1+exp{[(E^L(p,n)-\mu)/T]}} \; .
\eeq
The electron energy is restricted to magnitudes greater than $\mu$,
but the integrals in Eq.(\ref{7}) are insensitive to
$p_{min}$, so we take $p_{min}$=0 as in Ref\cite{fkmp03}.

 We agree with the estimate of Ref.\cite{fkmp03} for P(0). Note that if we
had used the free electron mass, $m_e$, in the Landau energies (Eq.(\ref{5}))
we would have obtained a much smaller value for P(0).
For B= $10^{16}$ G, $\mu=40$ MeV, $m_e^*$=4 MeV
and $T_{\nu-sphere}$ = 20 MeV, P(0) $\simeq$ 0.3. This is similar to our
estimate of P(n=0) $\simeq$ 0.4 at the surface of the protoneutron star
at about 10 seconds\cite{hjk07}. Therefore our result for asymmetric
emissivity differs from that of Fuller et al \cite{fkmp03} mainly in that 
we have a much larger mixing angle, and a much smaller effective volume, since
the sterile neutrinos oscillate back to active neutrinos within the
neutrinosphere; and therefore our emission only takes place near the
surface of the neutrinoshere. However, in contrast with purely active
neutrino emission in which the opacity results in very small pulsar
kicks~\cite{lq98}, the sterile neutrinos have a much larger effective
volume, and can therefore produce much larger pulsar velocities

\subsection{Estimate of $V_{eff}$= Effective Volume for Emission}

  To estimate $V_{eff}$ we make use of the early study of opacity in about
the first 20s of the creation of a neutron star via a supernova 
collapse~\cite{ip82,bl86,bml81}, and a recent detailed study of neutrino 
mean free paths\cite{ss07}. Since the mean free path of the sterile neutrino is
determined by that of the standard neutrino to which it oscillates, $\lambda$,
we make use of studies of active neutrino mean free paths. First note
that the neutrino mean free path is given by 
\beq
\label{14}
          1/\lambda &=& \int \frac{d^3 p}{(2 \pi)^3} M_{fi} [1-n(q)]
\nonumber \\
        &&  \times (1+ e^{(\mu_\nu -E_\nu)/kT}) \;,
\eeq
where $M_{fi}$ is the weak matrix element and $n(q)$ is the Fermi distribution.
For the calculation of the sterile neutrino, for which $M_{fi}=0$,
one can use the value of $1/\lambda$ with a factor of $sin^2(2\theta)$ from
the matrix element and another such factor from the occupation probability.
From the results of previous authors, for T in the 10 to 20 Mev range and 
$\mu$ in the 20 to 40MeV range, we estimate that  $\lambda \simeq 1.0 cm$ 
This gives a range for the effective sterile neutrino mean free path
\beq
\label{15}
        \lambda_s &\simeq& 5.0 {\rm \; to\;} 250 {\rm \; cm}  \; .
\eeq 

For a neutrinosphere radius of 40 km, with $\lambda_s << R_\nu$ this gives us 
$V_{eff}= (4\pi/3)( R_\nu^3-(R_\nu-\lambda_s)^3) \simeq 4\pi R_\nu^2 
\lambda_s$.

  From Eq.({\ref{10}), $R_\nu$ = 40 km, and $\lambda$=1.0 cm, 
\beq
\label{16}
   p_{ns}&=& M_{ns} v_{ns} \simeq \frac{0.67 \times 10^{25}}{ sin^2(2 \theta)}
 T_9^7 gm \frac{cm}{s} \; ,
\eeq
with $T_9 = \frac{T}{10^9 K}$.
Taking the mass of the neutron star to equal the mass of our sun, $M_{ns}
= 2 \times 10^{33}$ gm, we obtain for the velocity of the neutron star
\beq
\label{17}
        v_{ns} &\simeq& 3.35 \times 10^{-7} (\frac{T}{10^{10} K})^7 
\frac{1}{sin^2(2\theta)} \frac{km}{s} \; ,
\eeq.

For example, for T=10 Mev =$1.16\times 10^{11}$ kK,
\beq
\label{18}
             v_{ns} &\simeq& 47.3 \frac{km}{s} {\rm \; to \; }
2,370 \frac{km}{s} \; ,
\eeq
which means that sterile neutrino emission could account for the large
pulsar kick with the parameters extracted from Refs~\cite{ms07,s07}. If we
use the physical parameters that give Eq.(\ref{18}) for electron neutrinos,
we obtain a pulsar velocity of $v_{ns}$ = 95 km/s, which is consistent
with previous predictions by several authors.

  It should be noted that the study of the MiniBooNE and LSND results are
in progress, and the mixing angles that result could be much different from
those which we have used, changing our results. Preliminary data from the
MiniBooNE/Minos experiment\cite{minos08} is consistent with the MiniBooNE
results\cite{mini}. We also once more point out that Ref~\cite{mm08} questions
the accuracy of the peramaters extracted by Ref~\cite{ms07}.

\newpage

\section{Conclusions}

   Because of the strong magnetic fields in protoneutron stars and the
associated neutrinosphere, the electrons produced in the URCA processes
that dominate neutrino production in the first 10 seconds have a sizable 
probability, P(0), to be in the lowest (n=0) Landau level. This leads to 
asymmetric neutrino momentum. With the mixing angles found in 
Refs\cite{ms07,s07}, we find that the sterile neutrinos produced during 
this period for high luminosity pulsars can give the pulsars velocities of 
greater than 1000 km/s, as observed, similar to predictions based on sterile 
neutrinos as dark matter\cite{fkmp03}. Although the parameters found in the 
analyis\cite{ms07,s07} of the MiniBooNE\cite{mini}and LSND data is still in 
question, current analysis of the FermiLab experiment\cite{minos08} seems 
to be in agreement with MiniBooNE\cite{mini}.

   There is a strong correlation of the pulsar velocity with temperature, T.
Since it is difficult to determine T accurately, it is difficult for us to
predict the velocity of a pulsar whose kick arises from sterile neutrino
emission. On the other hand, if the pulsar kick arises from the asymmetric
emission of active neutrinos produced by the modified URCA processes after
10 seconds, also proportional to P(0)\cite{hjk07}, then T can be determined
by an accurate measurement of the neutrinos from the supernova. Therefore,
in future years, with much more accurate neutrino detectors, one could predict 
the velocity of the resulting pulsar. Unfortunately, the energy of emitted 
sterile neutrinos cannot be measured. From our results in the present paper
and those in Ref\cite{hjk07}, high luminosity pulsars receive a large kick 
both from sterile neutrinos in the first ten seconds and standard neutrinos 
in the second ten seconds.

This work was supported in part by DOE contracts W-7405-ENG-36 and 
DE-FG02-97ER41014. The authors thank Terry Goldman; and William Louis, 
Gerald Garvey and other LANL members of the MiniBooNE Collaboration for 
helpful discussions.

\end{document}